# Flavor Democracy in Particle Physics

## Saleh Sultansoy

*GaziUniversity, Dept. of Physics, 06500 Teknikokullar, Ankara, Turkey*
*Academy of Sciences, Institute of Physics, H. Cavid Av. 33, Baku, Azerbaijan*

**Abstract.** The flavor democracy hypothesis (or, in other words, democratic mass matrix approach) was introduced in seventies taking in mind three Standard Model (SM) families. Later, this idea was disfavored by the large value of the t-quark mass. In nineties the hypothesis was revisited assuming that extra SM families exist. According to flavor democracy the fourth SM family should exist and there are serious arguments disfavoring the fifth SM family. The fourth SM family quarks lead to essential enhancement of the Higgs boson production cross-section at hadron colliders and the Tevatron can discover the Higgs boson before the LHC, if it mass is between 140 and 200 GeV. Then, one can handle "massless" Dirac neutrinos without see-saw mechanism. Concerning BSM physics, flavor democracy leads to several consequences: $tan\beta \approx m_t /m_b \approx 40$ if there are three MSSM families; super-partner of the right-handed neutrino can be the LSP; relatively light E(6)-inspired isosinglet quark etc. Finally, flavor democracy may give opportunity to handle "massless" composite objects within preonic models.

## INTRODUCTION

It is known that the Standard Model (SM) with three fermion families well describes the large amount of particle physics phenomena [1]. However, there are a number of fundamental problems which do not have solutions in the framework of the SM: qurk-lepton symmetry and fermion's mass and mixing pattern, family replication and number of families, L-R symmetry breaking, electroweak scale etc. Then SM contains unacceptably large number of arbitrary parameters even in the three family case: 19 in the absence of right neutrinos (and Majorana mass terms for left neutrinos), 26 if neutrinos are Dirac particles and more than 30 if neutrinos are Majorana particles. Moreover, the number of "elementary particles", which is equal to 61 in three families case (36 quarks and anti-quarks, 12 leptons and anti-leptons, 12 gauge bosons and 1 Higgs boson), reminds the Mendeleyev Table. Four decades ago similar situation for hadrons (mesons and baryons) led to the quark model!

For these reasons, physicists propose a lot of different extensions of the Standard Model, most part of which predict a rich spectrum of new particles and/or interactions at TeV scale (see *e. g.* reviews [2, 3] and ref's therein). These extensions can be grouped in two classes, namely standard and radical ones.

Standard extensions remain in the framework of gauge theories with spontaneously broken symmetry and include: enlargement of Higgs sector, enrichment of fermion sector, introducing of new gauge symmetries etc. Radical extensions include: compositeness (preonic → pre-preonic models), SUSY (MSSM→SUGRA), extra space-time dimensions etc.

On the other hand, some points can be clarified from the basics of the SM itself. For example, existence of the right-handed components of neutrinos should be considered as the consequence of the quark-lepton symmetry: $\nu_R$'s are counterparts of the right-handed components of the up-type quarks. Therefore, observation of the neutrino oscillations does not "require new, beyond the Standard Model (BSM), physics", it is quite natural feature of the SM (assumption of massless neutrinos is the "relic" of pre-SM era)… Another example is the Flavor Democracy, which favors the existence of extra SM families. In the similar manner, this does not mean "BSM physics", we deal with the SM with four (or more) families.

It is known that the number of fermion families is not fixed by the Standard Model. Asymptotic freedom of the QCD suggests that this number is less than nine. Concerning the leptonic sector, the LEP data determine the number of light neutrinos to be N = $2.994 \pm 0.012$ [1]. Direct searches for the new leptons and quarks led to the following lower bounds on their masses [1]: $m_{ld} > 100.8$ GeV; $m_{vd} > 45$ GeV (Dirac type) and $m_{vd} > 39.5$ GeV (Majorana type) for stable neutrinos; $m_{vd} > 90.3$ GeV (Dirac type) and $m_{vd} > 80.5$ GeV (Majorana type) for unstable neutrinos; $m_{dd} > 199$ GeV (neutral current decays), $m_{dd} > 128$ GeV (charged current decays).

# FLAVOR DEMOCRACY AND THE STANDARD MODEL

Thirty years ago the *flavor democracy* (or democratic mass matrix) was proposed [4] in order to solve some problems of the Standard Model. However, in the three SM families case this approach leads to a number of unacceptable predictions, such as a low value of t-quark mass etc. On the other hand, flavor democracy seems very natural in the framework of SM and problems disappear if the fourth family is introduced [5-7] (see, also, review [8]).

The precision electroweak data does not exclude the fourth SM family, moreover, even a fifth or sixth SM family is allowed provided that the masses of their neutrinos are about 50 GeV [9, 10].

## Flavor Democracy

It is useful to introduce three different bases (see ref's [8] and [11] for details):
- Standard Model basis $\{f^0\}$,
- Mass basis $\{f^m\}$ and
- Weak basis $\{f^w\}$.

Usually the first and third bases are identified and this leads to lose of important hints. Three bases approach may be useful both for SM and BSM.

According to the three-family SM, before the spontaneous symmetry breaking quarks are grouped into following SU(2)×U(1) multiplets:

$$\begin{pmatrix} u_L^0 \\ d_L^0 \end{pmatrix}, u_R^0, d_R^0; \begin{pmatrix} c_L^0 \\ s_L^0 \end{pmatrix}, c_R^0, s_R^0; \begin{pmatrix} t_L^0 \\ b_L^0 \end{pmatrix}, t_R^0, b_R^0$$

In the one-family case all bases are equal and, for example, d-quark mass is obtained due to the Yukawa interaction

$$L_Y^{(d)} = a_d(\overline{u}_L \ \overline{d}_L)\begin{pmatrix} \varphi^+ \\ \varphi^0 \end{pmatrix} d_R + h.c. \Rightarrow L_m^{(d)} = m_d \overline{d}d$$

where $m_d = a_d\eta/\sqrt{2}$ and $\eta = <\varphi^0> = 246$ GeV. In the same manner, $m_u = a_u\eta/\sqrt{2}$, $m_e = a_e\eta/\sqrt{2}$ and $m_{ve} =$

$a_{ve}\eta/\sqrt{2}$ (if the neutrino is a Dirac particle). In the n-family case

$$L_Y^{(d)} = \sum_{i,j=1}^{n} a_{ij}^d(\overline{u}_{Li}^0 \ \overline{d}_{Li}^0)\begin{pmatrix} \varphi^+ \\ \varphi^0 \end{pmatrix} d_{Rj}^0 \Rightarrow L_m^{(d)} = \sum_{i,j=1}^{n} m_{ij}^d \overline{d}_i^0 d_j^0$$

where $d_1^0$ denotes $d^0$, $d_2^0$ denotes $s^0$ etc., and $m_{ij}^d = a_{ij}^d\eta/\sqrt{2}$. The diagonalization of the mass matrix of each type fermions, or in other words transition from SM basis to mass basis, is performed by well-known bi-unitary transformation:

$$d_{iL}^m = (U_L^d)_{ik} d_{kL}^0, \quad d_{jR}^m = (U_L^d)_{jl} d_{lR}^0$$

where superscripts 0 and m represent SM and mass bases, respectively. In this context the well-known CKM matrix is defined as $U_{CKM} = U_L^u(U_L^d)^+$ and contains 3 (6) observable mixing angles and 1 (3) observable CP-violating phases in the case of three (four) SM families. The weak basis is determined by the following transformation: $d_i^w = (U_{CKM})_{ij} d_j^m$.

*First Assumption:* Before the spontaneous symmetry breaking, all (down-type) quarks are massless and there are no differences between $d^0$, $s^0$, $b^0$ etc. In other words, fermions with the same quantum numbers are indistinguishable. This leads us to the first FD assumption; namely, Yukawa couplings are equal within each type of fermions: $a_{ij}^d \cong a^d$, $a_{ij}^u \cong a^u$, $a_{ij}^l \cong a^l$, $a_{ij}^v \cong a^v$. The first assumption results in n-1 massless particles and one massive particle with $m = na^F\eta/\sqrt{2}$ (F=u,d,l,v) for each type of the SM fermions.

*Second Assumption:* Because there is only one Higgs doublet which gives Dirac masses to all four types of fermions (up quarks, down quarks, charged leptons and neutrinos), it seems natural to make the second assumption; namely, Yukawa couplings for different types of fermions should be nearly equal: $a^d \cong a^u \cong a^l \cong a^v \cong a$. In the three-family case flavor democracy predicts $m_b \approx m_t \approx m_\tau \approx m_{v\tau} \approx 3a\eta/\sqrt{2}$. Considering the actual mass values of the third SM family fermions ($m_{v\tau} << m_\tau < m_b << m_t$), <u>the first and second assumptions lead to the statement that according to the flavor democracy the fourth SM family should exist.</u>

*Third Assumption:* It seems natural to put a equal to SU(2) gauge constant $g_W$. In this case $m_4 = 2\sqrt{2} g_W \eta \approx 450$ GeV. If a = 1, then $m_4 \approx 700$ GeV, which is close to the upper limit on heavy quark masses following from partial-wave unitarity at high energies

[12]. An ultimate upper limit for fourth family fermions is 2.5 TeV, which corresponds to $a^2/4\pi = 1$. The situation may be quite different in neutrino sector if their masses have Majorana nature (for details, see [11]). Let us mention that Majorana neutrinos relax the restriction on the number of extra families coming from the precision electroweak data because their give negative contribution to parameter S.

In terms of the mass matrix, the above arguments mean

$$M^0 = \frac{a\eta}{\sqrt{2}}\begin{pmatrix} 1 & 1 & 1 & 1 \\ 1 & 1 & 1 & 1 \\ 1 & 1 & 1 & 1 \\ 1 & 1 & 1 & 1 \end{pmatrix} \Rightarrow M^m = \frac{4a\eta}{\sqrt{2}}\begin{pmatrix} 0 & 0 & 0 & 0 \\ 0 & 0 & 0 & 0 \\ 0 & 0 & 0 & 0 \\ 0 & 0 & 0 & 1 \end{pmatrix}.$$

Therefore, the fourth family fermions are almost degenerate, in good agreement with experimental value $\rho = 0.9998 \pm 0.0008$ [1]. The masses of the first three SM family fermions, as well as observable interfamily mixings, are generated due to the small deviations from the full flavor democracy [11, 13, 14]. The parametrization proposed in [11] gives the values of the fundamental fermion masses and at the same time predicts the values of the quark and the lepton CKM matrices which are in good agreement with the experimental data. In principle, flavor democracy provides the possibility to obtain the small masses for the first three neutrino species without the see-saw mechanism [15].

*Arguments Against the Fifth SM Family:* The first argument disfavoring the fifth SM family is the large value of $m_t \approx 175$ GeV. Indeed, partial-wave unitarity leads to $m_Q \leq 700$ GeV $\approx 4m_t$ and in general we expect $m_t \ll m_4 \ll m_5$. Then, neutrino counting at LEP results in fact that there are only three "light" ($2m_v < m_Z$) non-sterile neutrinos, whereas in the case of five SM families four "light" neutrinos are expected. Finally, the degenerate fifth family is excluded at more than $5\sigma$ level by precision electroweak data.

## A Search for the Fourth SM Family

The fourth SM family quarks will be copiously produced at the LHC via gluon-gluon fusion (see [16] and references therein). The expected cross section is about 10 (0.25) pb for a quark mass of 400 (800) GeV. The fourth generation up-type quark, $u_4$, would predominantly decay via $u_4 \rightarrow Wb$, therefore, the expected event topologies are similar to those for t-quark pair production. The best channel for observing will be [17]:

$$gg \rightarrow u_4\bar{u}_4 \rightarrow WWb\bar{b} \rightarrow (l\nu)(JJ)b\bar{b}$$

where one W decays leptonically and the other hadronically. The mass resolution is estimated to be 20 (40) GeV for $m_{u4}$ = 300 (700) GeV. The small interfamily mixings [11, 14] lead to the formation of the fourth family quarkonia [16, 18]. The most promising candidate for the LHC is the pseudo-scalar quarkonium state, $\eta_4$, which will be produced resonantly resonantly via gluon-gluon fusion. Especially, the decay channel $\eta_4 \rightarrow ZH$ is the matter of interest [19].

The FNAL Tevatron Run II can observe $u_4$ and $d_4$ quarks if there is an anomalous interaction with enough strength between the fourth family quarks and known quarks [20-22].

The fourth family leptons will clearly manifest themselves at the future lepton and photon colliders [11, 23]. Also, the number of different fourth family quarkonium states can be produced resonantly at lepton and photon machines. Moreover, in difference from the LHC, states formed by up and down type quarks can be investigated separately even if their mass difference is small.

## The Higgs Boson

The fourth SM family leads to an essential increase (~8 times) of the Higgs boson production cross-section at hadron colliders and this can give the Tevatron experiments (CDF and D0) opportunity to discover the Higgs boson before the LHC, if its mass is between 140 and 200 GeV (see [24] and ref's therein). Both D0 and CDF Collaborations are looking for this opportunity [25, 26]. Already now, the results are placing constraints on the SM with four or more fermion families [27].

Concerning the LHC, it will be able to cover the whole region via the golden mode $H \rightarrow ZZ \rightarrow \ell\ell\ell\ell$ and detect the Higgs signal during the first year of operation if the fourth SM family exists [28-30].

# FLAVOR DEMOCRACY AND THE BSM PHYSICS

Below we give short remarks on two extensions of the SM, namely, MSSM and isosinglet quarks (for more details, see [31] and [8], respectively).

## Flavor Democracy and the MSSM

The huge number of free parameters [2, 31-33] in the three families MSSM leads to consideration of some simplified versions, such as the constrained

MSSM (see [34] and ref's therein). In general, these simplifications ignore interfamily mixings and existence of right-handed neutrinos (and consequently their super-partners). As the result one avoid possible conflicts with experimental data on flavor violating processes, but at the same time we also lose very interesting possible phenomenology.

*Flavor Democracy and tanβ:* According to the first assumption, in the framework of the three families MSSM, the masses of t- and b-quarks are as follows: $m_t = 3\lambda_u v_u$ and $m_b = 3\lambda_d v_d$ (for notations see [31]). Application of the second assumption, namely $\lambda_u \approx \lambda_d$, immediately leads to the relation $\tan\beta = v_u/v_d \approx m_t/m_b$. With $m_t \approx 175$ GeV and $m_b \approx 4.5$ GeV we obtain $\tan\beta \approx 40$. More conservatively, flavor democracy favors the region of $30 < \tan\beta < 50$ and lower values can be interpreted as an indication of the fourth MSSM family.

*RS-LSP Scenario:* Flavor Democracy favors the "right" sneutrinos (super-partners of the right-handed neutrinos) as the lightest supersymmetric particles. The RS-LSP scenario [31, 35] should be considered as a serious alternative to the neutralino-LSP scenario. Obviously, in the first case decay chains of the supersymmetric particles drastically differ from those of the second case (see [35, 36] and ref's therein). In this scenario, $\tilde{v}_R$ can be a viable candidate of cold dark matter [37].

## Flavor Democracy and Isosinglet Quarks

Another way to explain the relation $m_b << m_t$ is the introduction of new isosinglet down-type quarks (an example is the $E_6$-inspired [38, 39] extension of the SM fermion sector). This scenario leads to very interest predictions for the LHC [40, 41].

## ACKNOWLEDGMENTS

It is pleasure to thank the organizers of BPU-6 for the nice Conference and hospitality. I am grateful to my co-authors, especially, Engin ARIK, Orhan CAKIR, Ayla CELIKEL, Serkant Ali CETİN, Abbas Kenan CIFTCI, Rena CIFTCI, Rashid MEHDIYEV, Gokhan UNEL and Metin YILMAZ for fruitful collaboration.